\documentclass[aps,prx, reprint,longbibliography, nofootinbib,superscriptaddress]{revtex4-1}

\usepackage{graphicx,amsmath,amssymb,color,amsfonts,bm,physics}
\usepackage{mathtools,lipsum}
\usepackage[dvipsnames]{xcolor}
\usepackage[misc]{ifsym}

\usepackage[colorlinks=true]{hyperref}  
\hypersetup{
    bookmarks=true,         
    unicode=false,          
    pdftoolbar=true,        
    pdfmenubar=true,        
    pdffitwindow=false,     
    pdfstartview={FitH},    
    pdftitle={},    
    pdfauthor={},     
    pdfsubject={},   
    pdfcreator={},   
    pdfproducer={}, 
    pdfkeywords={} {} {}, 
    pdfnewwindow=true,      
    colorlinks=true,       
    linkcolor=blue, 
    citecolor=blue,        
    filecolor=magenta,      
    urlcolor=blue,           
        breaklinks=true
} 

\newcommand{\be}{\begin{equation}}
\newcommand{\ee}{\end{equation}}
\newcommand{\bea}{\begin{eqnarray}}
\newcommand{\eea}{\end{eqnarray}}

\definecolor{db}{rgb}{0,0,0.4}

\begin{document}

\title{Late-time ensembles of quantum states in quantum chaotic systems}

\author{Souradeep Ghosh} 
\affiliation{Department of Physics \& Astronomy, Texas A\&M University, College Station, TX 77843}

\author{Christopher M. Langlett}
\affiliation{Department of Physics \& Astronomy, Texas A\&M University, College Station, TX 77843}

\author{Nicholas Hunter-Jones}
\affiliation{Department of Physics and Department of Computer Science, University of Texas at Austin, Austin, TX 78712}

\author{Joaquin F. Rodriguez-Nieva}
\affiliation{Department of Physics \& Astronomy, Texas A\&M University, College Station, TX 77843}

\begin{abstract} 
Initially unentangled states undergoing quantum chaotic dynamics are expected to evolve into featureless states at late times. While this expectation holds true on an average, coarse-grained level, it is unclear if this expectation applies to higher statistical moments, as symmetries present in physical systems constrain the exploration of Hilbert space. Here we study the universal structure of late-time ensembles obtained from unitary dynamics in quantum chaotic systems with symmetries, such as charge or energy conservation. We find that while quantum states do not ergodically explore the entire Hilbert space at late times, the late-time ensemble typically becomes indistinguishable from Haar-random states in the thermodynamic limit and at the level of finite statistical moments. Importantly, our results apply to initial states easy to prepare in ongoing experiments---specifically, product states---that lie in the middle of the spectrum of quantum chaotic systems. We show that typically these states not only exhibit the same late-time ensemble average as Haar-random states, but also the same state-to-state fluctuations and higher statistical moments. In other words, there is no measurement---whether local or nonlocal---that one can do at the level of finite statistical moments to tell that states are not exploring the entire Hilbert space. Interestingly, within the class of low-entanglement initial states, we also find atypical initial conditions in the middle of the spectrum of Hamiltonians known to be `maximally chaotic'. Such atypical initial states have smaller variance of the symmetry operator relative to that of Haar random states, and evolve into non-universal ensembles that can be distinguished from the Haar ensemble by simple measurements or subsystem properties. In the limiting case of initial states with negligible variance of the symmetry operator (e.g., states with fixed particle number or energy eigenstates), the late-time ensemble has universal behavior captured by constrained random state ensembles. On the one hand, our results reveal that an extremely high level of quantum state randomness can typically be achieved even when dynamics is constrained by symmetries; on the other hand, our results show that one can still achieve strongly non-ergodic dynamics at `infinite temperature' within quantum chaotic regimes.
\end{abstract}

\maketitle

\section{Introduction}

The notions of chaos and ergodicity are the two main pillars in the theory of statistical mechanics. In classical many-body systems, chaotic dynamics resulting from nonlinear evolution leads to repulsion between classical trajectories and ergodic exploration of phase space. This allows us to replace temporal averages by ensemble averages over all states compatible with macroscopic constraints, such as total energy or particle number. Although defining notions of chaos and ergodicity in quantum regimes has been a long-standing challenge~\cite{1991PRL_Deutsch,1994PRE_Srednicki,1999JPA_Srednicki,2008Nature_Rigol,2016AnnPhys_ETHreview,2018RPP_Deutsch} due to the unitary nature of quantum evolution, in quantum statistical mechanics we draw from our classical intuition to replace temporal averages at late times with averages over all states compatible with the macroscopic constraints~\cite{2012PRL_thermlizationeisert,2016RPP_thermalizationreview,2025PhysRep_foundationsofstatmech}. In energy conserving systems, the Gibbs ensemble $\rho = e^{-\beta H}/{\rm Tr}[e^{-\beta H}]$ results from finding the most entropic ensemble compatible with the total energy of the system, with the diagonal structure arising from the random phase approximation, which accounts for the effect of quantum state averaging. Thus, a midspectrum initial condition gives rise to a late-time ensemble where the temporal average of an observable is equal to its average over all states in Hilbert space. 

However, when describing the late-time behavior of quantum states, there are three subtleties that need to be taken into consideration. First, in physical quantum systems, quantum states do not explore the entire state space even in the limit of infinite time, as symmetries typically present in these systems impose constraints on their dynamics. For instance, quantum states evolving under Hamiltonian (or energy conserving) dynamics, $|\Psi\rangle = \sum_{n=1}^D e^{-iE_n t}\langle n | \Psi_0\rangle |n\rangle$, have $D={\rm dim}[{\cal H}]$ constraints (where $\cal H$ is our Hilbert space) defined by the projection of the initial condition $|\Psi_0\rangle$ on the eigenstate basis $|n\rangle$. In this case, quantum evolution is constrained to a `high-dimensional torus' with fixed amplitudes $|\langle n | \Psi_0 \rangle |$, rather than over the full Hilbert space. Thus, a relevant question to ask is whether the lack of Hilbert space ergodicity has any {\it measurable} fingerprint in the late-time behavior of quantum states beyond its quantum state average, which is known to be accurately described by the Gibbs ensemble\cite{2025PhysRep_foundationsofstatmech,2010EPJ_vonneumann1929}.

Second, the late-time behavior of quantum states depends on the choice of the initial condition $|\Psi_0\rangle$, which determines the statistical properties of quantum states at late times from the onset of dynamics. For this reason, it is unclear whether it is possible to make {\it universal} statements about the statistical properties of quantum state at late times that are independent of $|\Psi_0\rangle$, particularly at the level of higher statistical moments, and beyond the average behavior already captured by the Gibbs ensemble. In addition, not all initial conditions are equally accessible---low-entanglement states, such as unentangled product states, are typically the easiest to prepare in many experimental platforms. This raises the question of whether this restricted class of initial states exhibit distinct late-time properties compared to more generic initial states.

Third, the timescales in which states explore uniformly the entire accessible Hilbert space are extremely long. In particular, temporal averages need to be performed over timescales proportional to the accessible Hilbert space to fully capture all the statistical moments of the late-time ensemble. That is, a midspectrum initial condition will take exponentially long times to explore the space of states. Such timescales are beyond the reach of existing experimental platforms. Only in the case of arbitrarily long sampling times, the resulting ensemble of states can be shown \cite{PhysRevX.14.041051} to be equivalent to the {\it random phase ensemble}~\cite{2009PRE_latetimeensemble,2012PRA_randomphase}, defined as the ensemble of states spanned by $|\Psi\rangle = \sum_{n=1}^D e^{i\theta_n}|\langle n | \Psi_0\rangle ||n\rangle$, where $\theta_n$ are random phases. This result is true so long as the Hamiltonian satisfies a no $k$-th resonance condition to all $k$-orders. 

These questions are very relevant for today's experiments with programmable quantum platforms, such as systems comprised of superconducting qubits~\cite{2019Nature_quantumsuppremacy,2024Science_googlekpz,2021Science_googlescrambling}, Rydberg atoms~\cite{2017Nature_51atomsimulator,2021Nature_Rydberg256,2021Nature_RydbertAFM}, and trapped ions~\cite{2017Nature_monroe,2021RMP_trappedtions}. These systems can retrieve extremely detailed statistical information about quantum states through their ability to take microscopically-resolved measurements of individual degrees of freedom, and their ability to evolve states with high precision and repetition rate. From such measurements one can retrieve the full distribution of outcome probabilities of a microscopic  observable~\cite{2023Nature_emergentdesign,2023PRL_fullcountingstatistics,2024PRB_fullcountingstatistics} or the subsystem entanglement entropy~\cite{2019Science_EEmbl,2018PRL_zollerrenyientropy,2019Science_renyiions,2023NatRev_randommeasurement}, both of which are sensitive to the fine-grained structure of quantum states. 
These capabilities are qualitative distinct from more traditional thermodynamic or transport probes, which measure coarse-grained observables (i.e., long wavelength and low frequency) such that higher statistical moments are effectively averaged out.  

These modern experimental capabilities have sparked a surge of theoretical activity aimed at understanding chaos and ergodicity beyond the average coarse-grained behavior of quantum states in chaotic systems. In particular, our current understanding of quantum chaos and ergodicity is rooted in the random matrix theory (RMT) behavior of eigensystem properties, such as the behavior of the level-spacing statistics~\cite{Atas2013,Atas2013a,2007PRB_ratiofactor} or the spectral form factor~\cite{2017JHEP_bhrmt,2017JHEP_chaosrmt,2018PRL_exactsff,Kos2018,2018PRX_qcdiagrammatics,Roy2020}, and the volume-law behavior of eigenstate entanglement entropy~(EE)~\cite{Murthy2019,2019PRE_lugrover,2017PRL_EEchaotic,2023PRE_averageEE, 2022PRX_eereview}. Recent efforts have aimed to go beyond these coarse features to capture `fine-grained' features, such as how spatial locality is imprinted in the structure of quantum states~\cite{Huang2021,2022PRE_deviationsfromETH,2024PRX_quantifyingchaos,2024arxiv_EEU1}, and how this structure gives rise to fine features in the level spacing statistics, the matrix elements of local observables, or between eigenstates~\cite{2019PRL_chalker,2019PRE_kurchan,2020PRE_beyondETH,Garratt2021,2022PRL_dymarsky,2021PRE_eth_otocs}. Other works ~\cite{2023PRX_cotleremergentdesign,2022PRL_exactdesigns,2022Quantum_designs,2023PRA_freedeepthermalization,2023PRL_completeergodicity,2024arxiv_completeergodicity2} have aimed to quantify randomness of quantum state ensembles at the level of higher statistical moments using $k$-designs~\cite{HunterJones2019,roberts2017chaos}. Many other statistical approaches are being developed to capture higher-order fluctuations of matrix elements of local observables~\cite{2023arxiv_fava,2020PRA_higherorderETH,2022PRL_pappalardi,2023arxiv_papalardi}. 
Here, we focus on the question of universality in the late-time ensemble at the level of higher statistical moments, examining how its behavior depends on the choice of initial condition and how the lack of ergodicity affects the quantum state statistics beyond their average, thermal behavior. Specifically, we consider the late-time dynamics of initial product states (i.e., unentangled states) lying in the {\it middle of the spectrum} of a quantum chaotic system, and assume that states can be sampled at arbitrarily long times. Although such states are generally expected to evolve into featureless states, here we find much richer dynamical behaviors that arise when looking at the fine-grained structure of quantum states. In particular, for quantum chaotic systems with spatial locality and energy conservation as their key features, we show the existence of two limiting late-time regimes with distinct universal properties that depend on the initial condition---both late-time regimes agree on the usual indicators of quantum chaos, including `volume law' entanglement entropy and average behavior described by the Gibbs ensemble, but exhibit qualitatively distinct behaviors at the level of higher statistical moments. They key results of the present work are as follows. 

\subsection{Summary of key results}
\label{sec:results}

We analyze the statistical properties of quantum state ensembles at {\it late times} generated by a unitary evolution operator $\hat{U}_t$. The late-time ensemble is defined as 
\be
\Phi_{T\rightarrow\infty}=\left\{|\Psi_i\rangle = \hat{U}_{t_i} |\Psi_0\rangle, \, t_i\in [T,\infty]\right\},
\ee 
where the states $|\Psi_i\rangle$ are drawn randomly at times $t_i$ after a long initial time $T\gg 1$ has passed. We focus on quantum chaotic dynamics constrained by a symmetry $\hat Q$ describing a local conserved quantity, such as magnetization, particle number or energy. In addition, we focus on low-entanglement initial conditions $|\Psi_0\rangle$ typically prepared in experiments, specifically product state initial conditions. Specifically, we ask whether the constraint introduced by $\hat Q$ gives rise to distinguishable features relative to pure random states that can be measured in experiments, e.g., through measurements of subsystem properties or simple observables. 

\begin{figure}
    \includegraphics[width=\columnwidth]{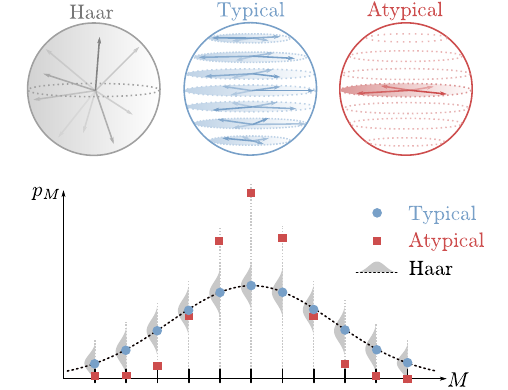}
    \caption{The late-time behavior of an initial state $|\Psi_0\rangle$ can be classified by its distribution of outcome probabilities of the symmetry operator $Q = \sum_n Q_n|n\rangle\langle n|$ relative to that produced by the ensemble of Haar random states ($Q_n$ and $\ket{n}$ and the eigenvalues and eigenstates of $Q$). A typical initial state (circles) has the same distribution of outcome probabilities as Haar random states (shaded gray distributions). In contrast, an atypical initial state (squares) has a distribution which has smaller than typical fluctuations of the symmetry operator. Schematics for typical and atypical states are shown for a magnetic charge $Q = \sum_i Z_i$ with quantum number $M$.}
    \label{fig:schematics}
\end{figure}

We find that the late-time behavior of $|\Psi_0\rangle$ can be characterized in terms of the distribution of outcome probabilities of the symmetry operator $\hat{Q}$ (Fig.~\ref{fig:schematics}). First, consider the spectral decomposition of the operator $\hat{Q}$, given as $\hat{Q} = \sum_n Q_n |n\rangle\langle n|$, where $\ket{n}$ are the eigenvectors of $\hat{Q}$ and $Q_n$ the corresponding eigenvalues. The initial state $|\Psi_0\rangle$ gives rise to a probability distribution $f(Q)$ which can be expressed in the eigenbasis of the operator $\hat Q$ as 
\be 
f(Q) = \sum_n |\langle n | \Psi_0\rangle|^2\delta(Q-Q_n).
\label{eq:f(O)}
\ee
All the moments of $f(Q)$ are conserved during the evolution, as each factor $f_n = |\langle n | \Psi_0 \rangle|^2$ remains constant. 
Similarly, the Haar ensemble $\Phi_{\rm Haar}$ of pure random states in the full Hilbert space $\cal H$ generates a reference distribution
\be
g(Q) = \sum_n  \langle |\langle n | \Psi\rangle|^2\rangle \,\delta(Q-Q_n),
\label{eq:g(O)}
\ee
where the outer brackets denote average over Haar random states, $\langle [\ldots]\rangle = \int \prod_\sigma d\Psi_\sigma d\bar{\Psi}_\sigma [\ldots]\delta(|\Psi|^2-1)$. 
Due to the randomness of $|\Psi\rangle$, each value of $g_n = \langle |\langle n | \Psi \rangle|^2\rangle=1/D$ is uniformly distributed, with $D = {\rm dim}[{\cal H}]$. 

We say that an initial product state $|\Psi_0\rangle$ is {\it typical} if it satisfies the condition
\be
f(Q) = g(Q).
\label{eq:typical}
\ee
This is equivalent to requiring that the moments of the operator $Q$—such as $\langle Q \rangle_{\Psi_0}$, $\langle Q^2 \rangle_{\Psi_0}$, and so on—evaluated for the initial state $|\Psi_0\rangle$, match those of Haar-random states. In the case of quantum chaotic Hamiltonians, it is not hard to satisfy the condition (\ref{eq:typical}), as product states typically spread uniformly across the eigenbasis of $\hat Q$. If the system exhibits additional symmetries, the condition (\ref{eq:typical}) needs to be generalized to the full joint distribution of all conserved quantities. 

For typical initial states, we find that the late-time ensemble and the Haar ensemble share exactly the same finite $k$-moments in the thermodynamic limit, even though the time-evolved states do not explore the entire Hilbert space. {\it In other words, there is no measurement---whether local or nonlocal---that one can do at the level of finite statistical moments to tell that states are not exploring the entire Hilbert space.} In the case of finite-sized systems, it is exponentially hard to distinguish the late-time ensemble from the Haar ensemble. For instance, when considering subsystem entropies, the late-time ensemble of a typical initial state has an average EE equal to the Page entropy~\cite{1993PRL_Page} up to, possibly, exponentially-small-in-subsystem corrections. 
The reason behind this behavior is that, although symmetries constrain evolution and not all states are explored at late times, product state initial conditions typically mix all symmetry sectors at the onset of quantum evolution and give rise to Haar-like moments at late times. We show that this behavior is robust across families of quantum chaotic Hamiltonians.

If the distributions $f(Q)$ and $g(Q)$ disagree (Fig.~\ref{fig:schematics}), i.e., the initial state is {\it atypical}, then the late-time ensemble is distinguishable from the Haar ensemble at the level of the first moment, and such differences can be detected from subsystem properties or from simple local measurements. For example, in systems with charge conservation, the atypicality condition can be diagnosed either directly by measuring the charge distribution at each site, or indirectly by detecting ${\cal O}(1)$ corrections (or fluctuations) to the Page entropy, as we discuss below. A main focus of the present work is when the first moment of $f(Q)$ and $g(Q)$ agree but higher moments do not. In this case, the initial state is a `midspectrum' initial states, because $\langle \Psi_0 |{\hat Q}|\Psi_0\rangle = {\rm Tr}[Q]$, but higher moments of $\hat Q$ are atypical. As we will see, this leads to deviations from pure random states which are nonuniversal and depend on the details of $f(Q)$. 
We show that atypical states are not rare, even when considering product state initial conditions, and can easily be found even in the middle of the spectrum of strongly chaotic Hamiltonians. 

In the limiting case of initial conditions with negligible variance of the symmetry operator, such as states with fixed particle number or energy eigenstates, the late-time ensemble acquires universal behavior described by constrained random state ensembles. In essence, such ensembles are Haar random states constrained to a specific sector of the symmetry operator, capturing its microcanonical properties. Such constrained ensembles exhibit lower entanglement entropy than pure random states, with ${\cal O}(1)$ corrections to the entanglement entropy and larger statistical fluctuations compared to those of the Haar ensemble. Physically, entanglement entropy below the Page value indicates that eigenstates contain more structure and information than Haar-random states. This additional structure was shown to arise the combination of spatial locality and energy conservation, both of which are imprinted on the structure of eigenstates~\cite{Huang2021,2024PRX_quantifyingchaos}. This behavior is qualitatively distinct from that of typical initial states, where spatial locality is erased by dynamics at late times, giving rise to Haar-like behavior.

The outline for the rest of the work is as follows. In Sec.~\ref{sec:proof},  we use a simple analytically-tractable model to prove that constrained ensembles that satisfy the typicality condition are indistinguishable from Haar random states at the level of finite $k$ moments and in the thermodynamic limit. In Sec.~\ref{sec:EE}, and in sections thereafter, we analyze the statistical properties of quantum state ensembles from the lens of the von Neumann entanglement entropy (extensions to distributions of different observables is discussed in the Appendix). Thus, we first review in Sec.~\ref{sec:EE} the relevant entanglement distributions for constrained and unconstrained RMT ensembles relevant to our work. In Sec.~\ref{sec:U(1)}, we show numerical results for random quantum circuits in the presence of a U(1) scalar charge. In Sec.~\ref{sec:Ham}, we extend the numerical results to local chaotic Hamiltonian systems, showing that the same conclusions obtained in Sec.~\ref{sec:proof} apply to more generic quantum chaotic systems. In Sec.~\ref{sec:discussion}, we summarize the main results and discuss their ramifications.

\section{Moments of constrained quantum state ensembles: exact results}
\label{sec:proof}

We now prove the main results of the present work in a simple and analytically-tractable model, namely, ensembles of pure random states in the presence of a U(1) constraint. We consider a chain of $L$ spin-1/2 degrees of freedom with total Hilbert space dimension $D = 2^L$. We factor the Hilbert space into $(L+1)$ magnetic sectors of the symmetry operator ${\hat Z} = \sum_i^L {\hat Z}_i$, with $M$ ($0 \le M \le L$) labelling the total number of spins down in a given sector. Each magnetic sector has Hilbert space dimension $D_M = {L \choose M}$, with $\sum_{M=0}^L D_M = D$. 

Let us now draw pure random states from the ensemble defined as
\be
|\Phi\rangle = \sum_{M=0}^L \sqrt{p_M} |\Phi_M\rangle,
\label{eq:phiM}
\ee
where $p_M = D_M/D$ is fixed, and $|\Phi_M\rangle \in {\cal H}(M)$ are normalized random states within the subspace with fixed $M$ particles, $|\Phi_M\rangle = \sum_{\alpha=1}^{D_M} \phi_{M,\alpha}|M,\alpha\rangle$, with $\sum_{\alpha = 1 }^{D_M}|\phi_{M,\alpha}|^2=1$ and $\alpha$ labeling the states within the sector $M$.  The states $|\Phi\rangle$ are properly normalized because of the condition $\sum_M p_M = 1$. Importantly, we note that states drawn from Eq.~(\ref{eq:phiM}) are typical, as pure random states drawn from the Haar ensemble also have projection $\langle\langle \Psi | \hat{P}_M | \Psi\rangle\rangle = p_M$ 
in each symmetry sector with quantum number $M$, where ${\hat P}_M$ denotes the projection operator on the magnetic sector $M$, ${\hat P}_M = \sum_{\alpha=1}^{D_M}|M,\alpha\rangle\langle M,\alpha|$.

Because the values of $p_M$ are fixed, the ensemble $\Phi$ in Eq.~(\ref{eq:phiM}) does not explore the entire Hilbert space. For instance, the states $\ket{\uparrow\uparrow\ldots\uparrow}$ and $\ket{\uparrow\downarrow\uparrow \ldots\uparrow}$ are not part of $\Phi$. Nevertheless, we will see that $\Phi$ and the Haar ensemble are indistinguishable for all finite $k$ moments in the thermodynamic limit. In the case of finite-sized systems, differences between finite moments are ${\cal O}(1/D)$ and, therefore, hard to measure. Before proving the previous statement, we first review some basics of ensemble averaging. 

\subsection{Ensemble of Haar Random States}

Let us consider pure random states in a Hilbert space of dimension $D$, $|\Psi\rangle = \sum_{n=1}^{D} \psi_n|n \rangle$. One can think of $|\Psi\rangle$ as a $D$-dimensional vector of randomly distributed Gaussian variables which are not independent as they must satisfy the normalization condition, $\sum_{n=1}^D |\psi_n|^2=1$. The $k$-th moment of the Haar ensemble is defined in terms of the $D^k\times D^k$ density matrix obtained by averaging $k$ copies of the state $|\Psi\rangle$:
\be
\rho_{}^{(k)} =  \big\langle{\, \underbrace{|\Psi\rangle \langle \Psi| \otimes \ldots \otimes|\Psi\rangle \langle \Psi|}_{k\,\,{\rm copies}}}\,\big\rangle.
\ee
For example, the first moment of the distribution is 
\be
\rho_{m,n}^{(k=1)} =\langle \psi_m\bar{\psi}_{n}\rangle = \frac{\delta_{m,n}}{D}.
\label{eq:Haarfirstmoment}
\ee
This equality can be obtained from taking ensemble average over the normalization condition $\sum_n \langle \psi_n\bar{\psi}_n\rangle = 1$, and noting that all the components of the wavefunction are equivalent, which results in each of the  wavefunction's component having variance $1/D$. In addition, ensemble average over distinct components gives zero, as each component is uncorrelated. 

The second moment of the Haar ensemble is 
\begin{align}
  \rho_{m_1m_2,n_1n_2}^{(k=2)} & = \langle \psi_{m_1}\bar{\psi}_{n_1}\psi_{m_2}\bar{\psi}_{n_2}\rangle \nonumber\\
  & = \frac{\delta_{m_1,n_1}\delta_{m_2,n_2}+\delta_{m_1,n_2}\delta_{m_2,n_1}}{D(D+1)}.
\label{eq:Haarsecondmoment}
\end{align}
Similar to the $k=1$ case, this result can be obtained from taking ensemble average of the square of the normalization condition $\sum_{n,n'}\langle \psi_{n}\bar{\psi}_{n}\psi_{n'}\bar{\psi}_{n'}\rangle =1$, which contains $D(D-1)$ equivalent terms with $n\neq n'$, and $D$ terms with $n = n'$ which contribute twice to the summation (as there are two ways to contract the Gaussian variables). This gives rise to a total of $D(D+1)$ equivalent terms that sum to 1.

More generally, one can prove \cite{2013arxiv_harrowmoments} that
\be
\rho_{\vec{m},\vec{n}}^{(k)} = \bigg\langle \prod_{i=1}^{k} \psi_{m_i}\bar{\psi}_{n_i}\bigg\rangle = \frac{\sum_{{\vec\sigma}\in P_k}\delta_{{\vec m},P_{\vec\sigma}({\vec n})}}{\prod_{j=0}^{k-1}(D+j)}\,,
\label{eq:haarmoments}
\ee
where $\vec m = (m_1 , \ldots, m_k)$ and $\vec n = (n_1,\ldots,n_k)$ denotes vectors labeling the entries of the $D^k \times D^k$ matrix in the basis $|n\rangle$, $P_{\vec\sigma}({\vec n})$ denotes permuting the indices of the vector $\vec n$ for all possible permutations $\vec\sigma$ of $k$ indices, and $\delta_{{\vec n},{\vec m}}$ is a short-hand notation for $\delta_{{\vec n},{\vec m}} = \delta_{{n_1},m_1}\delta_{n_2,m_2}\ldots\delta_{n_k,m_k}$. Equivalently, $k$-th moments of Haar random states are equal to the projector onto the symmetric subspace of $k$ copies of the Hilbert space.

From Eq.~(\ref{eq:haarmoments}) we see that, when we consider finite $k$-moments of the distribution in the thermodynamic limit $D \rightarrow\infty$, it is legitimate to replace Eq.~(\ref{eq:haarmoments}) by 
\be
\rho_{\vec{m},\vec{n}}^{(k)} = \bigg\langle \prod_{i=1}^{k} \psi_{m_i}\bar{\psi}_{n_i}\bigg\rangle \approx \frac{\sum_{{\vec\sigma}\in P_k}\delta_{{\vec m},P_{\vec\sigma}(\vec{n})}}{D^k},
\label{eq:haarmomentsthermo}
\ee
where subleading terms are neglected relative to the leading order contribution in $1/D$. This is exactly the same result that one would get for an unnormalized random vector with $D$ independent random Gaussian variables.

\subsection{Ensemble of Typical Random States}

Let us now compute the first few moments of the distribution $|\Phi\rangle$ in Eq.~(\ref{eq:phiM}).
Here we label states in the eigenbasis of the $\hat Z$ operator, $m = (M,\alpha)$, with $1\le \alpha \le D_M$. The first moment of the ensemble $\Phi$ is 
\begin{align}
  \rho_{(M\alpha),(N\beta)}^{(k=1)} & = \sqrt{p_Mp_N}\langle \phi_{M\alpha}\bar{\phi}_{N\beta} \rangle \nonumber\\
  & = p_M \frac{\delta_{M\alpha,N\beta}}{D_M} = \frac{\delta_{n,m}}{D},
  \label{eq:U1firstmoment}
\end{align}
where in the third equality we used $p_M = D_M/D$ combined with Eq.~\eqref{eq:Haarfirstmoment}
applied to pure random states within the symmetry sector with quantum number $M$. In the last equality in Eq.~\eqref{eq:U1firstmoment} we relabelled the states as $|n\rangle = |M\alpha\rangle$. Equation~\eqref{eq:U1firstmoment} agrees {\it exactly} with Eq.~\eqref{eq:Haarfirstmoment}, thus the ensemble of Haar random states and the ensemble of typical states have the same first moment.  

Calculation of the $k=2$ moment of the ensemble $\Phi$ is somewhat more complicated and needs to be separated into two different cases: (i) $M_1 \neq M_2$, and (ii) $M_1 = M_2$. For $M_1\neq M_2$, we find
\begin{align}
\rho_{{\vec M}{\vec\alpha},{\vec N}{\vec\beta}}^{(k=2)} & = \sqrt{p_{M_1}p_{N_1}p_{M_2}p_{N_2}}\langle \phi_{M_1\alpha_1}\bar{\phi}_{N_1\beta_1}\phi_{M_2\alpha_2}\bar{\phi}_{N_2\beta_2} \rangle  \nonumber\\
  & = p_{M_1}p_{M_2} \frac{\delta_{M_1\alpha_1,N_1\beta_1}\delta_{M_2\alpha_2,N_2\beta_2}+(N_1\beta_1\rightleftharpoons N_2\beta_2)}{D_{M_1}D_{M_2}} \nonumber\\ &= \frac{\delta_{m_1,n_1}\delta_{m_2,n_2}+\delta_{m_1,n_2}\delta_{m_2,n_1}}{D^2}.
\label{eq:U1secondmoment1}
\end{align}
For $M_1 = M_2$, we find 
\begin{align}
\rho_{{\vec M}{\vec\alpha},{\vec N}{\vec\beta}}^{(k=2)} & = p_{M_1}^2 \frac{\delta_{M_1\alpha_1,N_1\beta_1}\delta_{M_1\alpha_2,N_2\beta_2}+(N_1\beta_1\rightleftharpoons N_2\beta_2)}{D_{M_1}(D_{M_1}+1)} \nonumber\\
  & = \frac{\delta_{m_1,n_1}\delta_{m_2,n_2}+(n_1\rightleftharpoons n_2)}{D^2(1+D_{M_1}^{-1})}.
  \label{eq:U1secondmoment2}
\end{align}

Unlike the first moment, the second moment of the ensemble of typical states deviates from that of Haar-random states, Eq.~(\ref{eq:Haarfirstmoment}). To quantitatively estimate the difference  between the second moments, we first compute the difference $\delta\rho^{(k)}$ between the matrices $\rho_{\vec{m},{\vec n}}^{(k=2)}$ for Haar random states and typical random states, and then compute its trace distance $\Delta_{k}$. The trace distance of a matrix is defined as one half the sum of the absolute values of its eigenvalues. When the trace distance is computed on the difference between two density matrices, each of which has a trace equal to one, the minimum value of $\Delta_k$ is 0 (when the density matrices are equal) and its maximum value is 1. As shown in the Appendix, the trace distance $\Delta_{2}$ between ensembles is given by
\be
\Delta_{2} = \frac{1}{D}\left(1-\frac{1}{\sqrt{\pi L}}\right) + {\cal O}\left(\frac{1}{D^2}\right), 
\label{eq:Delta2}
\ee
which is exponentially small in system size, and goes to zero in the thermodynamic limit. 

For higher moments, $k>2$, we take a shortcut and start the calculation by assuming $D\rightarrow\infty$. In this case, we can use the analogue of Eq.~(\ref{eq:haarmomentsthermo}) for each symmetry sector $M$ and assume that each $D_M$-dimensional vector $\phi_{M,\alpha}$ has random and independently-distributed components with variance $1/D_M$ ({i.e.}, no normalization constraint). Then, the $k$-th moment of the ensemble $\Phi$ in the thermodynamic limit is 
\begin{align}
  \rho_{(\vec{M}\vec{\alpha}),(\vec{N}\vec{\beta})}^{(k)} & = \sqrt{p_{M_1}p_{N_1}\ldots p_{M_k}p_{N_k}}\Big\langle \prod_{i=1}^{k}\phi_{M_i\alpha_i}\bar{\phi}_{N_i\beta_i} \Big\rangle \nonumber\\
  & = p_{M_1}\ldots p_{M_k}\frac{\sum_{{\vec\sigma}\in P_k}\delta_{{\vec M}{\vec\alpha},P_{\vec\sigma}({\vec N}{\vec\beta})}}{D_{M_1}\ldots D_{M_k}} \nonumber\\
     & = \frac{\sum_{{\vec\sigma}\in P_k}\delta_{{\vec m},P_{\vec\sigma}({\vec n})}}{D^k}.
     \label{eq:ktypical}
\end{align}
In this case, Eq.~(\ref{eq:ktypical}) agrees {\it exactly} with Eq.~(\ref{eq:haarmomentsthermo}) for all finite $k$ in the thermodynamic limit. 

\subsection{Ensemble of Atypical Random States}

For the sake of completeness, let us now consider the microcanonical ensemble of pure random states constrained to the largest symmetry sector ($M_*=L/2$) of the $\hat Z$ operator, 
\be
|\Phi'\rangle = \sum_{\alpha=1}^{D_{M_*}}\phi_{M_*,\alpha}|M_*,\alpha\rangle.
\label{eq:atypical}
\ee
The first moment of the ensemble of $\Phi'$ is given by
\begin{align}
\rho_{M\alpha,N\beta}^{(k=1)} = \delta_{M,M_*}\langle \phi_{M,\alpha}\bar{\phi}_{N,\beta} \rangle = \delta_{M,M_*}\frac{\delta_{M\alpha,N\beta}}{D_{M_*}}, 
\end{align}
which already differs from the first moment of the Haar ensemble, Eq.~(\ref{eq:Haarfirstmoment}). 
In this case, the trace distance between the first moment of the Haar ensemble and the ensemble $\Phi'$ is given by 
\be
\Delta_{1}= \left( 1-\sqrt{\frac{2}{{\pi L}}}\right),
\ee
which is nearly the maximal distance between the two ensembles. As shown below, such differences are measurable from subsystem properties.

\section{Moments of entanglement entropy within ensembles}
\label{sec:EE}

In what follows, we present all numerical data in terms of the distribution of the von Neumann entanglement entropy (EE),
\be
S_A = -\text{Tr}[\rho_A \log \rho_A],
\label{eq:EE}
\ee
in order to characterize and quantify quantum state randomness at late times. In Eq.~(\ref{eq:EE}), the reduced density matrix of a given quantum state $|\Psi\rangle$ is defined as $\rho_A = {\rm Tr}_B[|\Psi \rangle\langle\Psi|]$, where the system is bipartitioned into two subsystems of sizes $L_A = f L$ and $L_B=(1-f)L$. In addition, we focus specifically on the half-system EE, i.e., $f=1/2$. There are several reasons for focusing on the statistical properties of half-system EE. First, the EE is a nonlinear function of $\rho_A$, making it a particularly sensitive probe of proximity to Haar randomness.  Second, the exact distribution of EE is known for various classes of constrained~\cite{2019PRD_BianchiDona,2022PRX_eereview} and unconstrained~\cite{2016PRE_entanglementdispersion,2017PRE_entanglementvariance_proof,2019PRD_BianchiDona} ensembles of pure random states, which we briefly review below. Third, analyzing the half-system EE of quantum state ensembles allows us to perform a worst-case scenario analysis: if its distribution is indistinguishable from that of Haar random states, then the same result holds for any smaller subsystems or finite Renyi entropies. This follows from the fact that as $L_A$ decreases, the subsystem $A$ is coupled to an increasingly larger bath, making $\rho_A$ effectively `more thermal'.

A drawback of focusing on the half-system EE is that it is challenging to measure experimentally. This difficulty arises because $S_A$ is a nonlinear function of $\rho_A$ and because obtaining $\rho_A$ requires measuring a number of degrees of freedom that scales exponentially with system size. For this reason, in the Appendix, we extend our analysis to show that the conclusions of this work also hold in experimentally accessible scenarios. In particular, we demonstrate that our results apply to the EE of smaller subsystems ($f<1/2$), as well as to other more easily measurable quantities, such as the second Renyi entropy.

We now briefly review the statistical behavior of EE in ensembles of pure random states, considering both unconstrained and constrained ensembles. Our focus will be on the {\it asymptotic} behavior of the first and second moments of the EE distribution for arbitrary $f$, while referring the reader to Refs.~\cite{2016PRE_entanglementdispersion,2017PRE_entanglementvariance_proof,2019PRD_BianchiDona,2022PRX_eereview} for exact analytical expressions.

\subsection{Entanglement patterns for the Haar ensemble}

In the absence of any structure, the distribution of subsystem entanglement entropy of pure random states drawn from $\Phi_{\rm Haar}$ depends only on subsystem dimensions through the parameters $f$ and $L$. Without loss of generality, we assume $f \le 1/2$. The average EE $\mu_{\rm H} = \langle S_A\rangle $ is given by 
\be
 \mu_{\rm H} \approx Lf\log d-\frac{1}{2}\delta_{f,1/2},
\label{eq:page}
\ee
which was first conjectured by Page~\cite{1993PRL_Page} and later proven analytically by others~\cite{2016PRE_entanglementdispersion,2017PRE_entanglementvariance_proof,2019PRD_BianchiDona}.
The first term in the RHS of Eq.~(\ref{eq:page}) is the volume-law term which scales with subsytem size $L_A = fL$, and the second term gives rise to the `half-qubit' shift correction for half subsystems.
The variance of EE for pure random states, $\sigma^{2}_{\rm H} \approx d^{-L(1+|1-2f|)}$, is exponentially small in subsystem size. This implies that the EE is typical and a single pure random state has the Page entropy in Eq.~(\ref{eq:page}). 

\subsection{Entanglement patterns for constrained ensembles: U(1) and energy conserving systems}
\label{sec:EEU1}

For systems with a scalar charge (such as magnetization or particle number) and a local Hilbert space dimension of $d=2$, it is convenient to think of $0 \leq M \leq L$ as an integer charge number, and each site able to accommodate a maximum of one charge only. This is exactly true for conservation of magnetization or particle number, but only an approximation in energy conserving systems. The Hilbert space ${\mathcal{H}(M)}$ of states with fixed charge $M$ decomposes as a direct sum of tensor products, 
\be
{\cal H}(M) = \bigoplus_{M_A } {\cal H}_A(M_A)\otimes{\cal H}_B(M-M_A),
\label{eq:u1factorization}
\ee
where $M_A$ is an integer number varying within the range ${\rm max}(0, M-L_B) \le M_A \le {\rm min}(M, L_A)$. The Hilbert space dimension of ${\cal H}_A(M_A)$ is $d_{A,M_A} = \binom{L_A}{M_A}$, the Hilbert space dimension of ${\cal H}_{\rm B}(M-M_A)$ is $d_{B,M-M_A}= \binom{L-L_A}{M-M_A}$, and the total Hilbert space dimension is $\sum_{M_A} d_{A,M_A} d_{B,M-M_A} =d_M =\binom{L}{M}$. A random state within a fixed charge sector $|\Phi_M\rangle \in {\cal H}(M)$ can be expressed as a superposition of orthonormal basis states, $|\Phi_M\rangle = \sum_{M_A}\phi_{\alpha,\beta}^{(M_A)}|M_A,\alpha\rangle\otimes|M-M_A,\beta\rangle$, 
with $\phi^{(M_A)}_{\alpha,\beta}$ uncorrelated random numbers up to normalization.
The index $\alpha$~($\beta$) labels the basis states in subsystem $A$~($B$) with a total charge $M_A$~($M-M_A$).

The reduced density matrix of subsystem $A$ is block diagonal, $\rho_{A}= {\rm Tr}_B[|\Phi_M\rangle\langle\Phi_M|]= \sum_{M_A} p_{M_A} \rho_{A|M_A}$, where $\rho_{A|M}$ denotes the block with $M_A$ particles, and the factors $p_{M_A}\ge 0$ are the (classical) probability distribution of finding $M_A$ particles within $A$. The EE can be written as $S(\rho_{A}) = \sum_{M_A} p_{M_A} S(\rho_{A|M_A})- p_{M_A} \log p_{M_A}$, where the second term on the RHS is the Shannon entropy of the number distribution $p_{M_A}$, which captures particle number correlations between the two subsystems, and the first term captures quantum correlations between configurations with a fixed particle number. 

The first few moments of the EE distribution produced by $|\Phi_M\rangle \in {\cal H}(M)$ were first computed in Ref.~\cite{2019PRD_BianchiDona}. The mean entanglement entropy for `mid-spectrum' states ($M/L=1/2$) in the asymptotic limit is given by 
\be
\mu_M = Lf\log(d)+\frac{f+\log(1-f)}{2}-\frac{1}{2}\delta_{f, 1/2}.
\label{eq:meanu1}
\ee
Interestingly, in addition to the volume-law term and the half-qubit shift, 
Eq.~(\ref{eq:meanu1}) also exhibits a finite shift in the mean EE entropy relative to the Haar result in Eq.(\ref{eq:Haarfirstmoment}). 
The variance of EE scales exponentially with system size, $\sigma_{M} \sim \sqrt{L}/d^{L}$, thus a typical pure random state in ${\cal H}(M)$ will have the EE in Eq.~(\ref{eq:meanu1}).
We emphasize that the differences between the average EE in Eqs.~(\ref{eq:page}) and (\ref{eq:meanu1}) are significant on the exponentially small scale set by $\sigma_{M}$. 

While the above analysis is exact for systems with a local scalar charge---where the charge is expressed as a sum of commuting single-site operators---in Refs.~\cite{2024PRX_quantifyingchaos,2024arxiv_EEU1} we argued that the EE distribution of the microcanonical eigenstate ensemble (i.e., the analogue of states with zero magnetization variance) in local Hamiltonian systems exhibits the same asymptotic behavior as in systems with a local scalar charge discussed above. In particular, while the Hilbert space of local Hamiltonian systems cannot be factored as in Eq.~(\ref{eq:u1factorization})---since the energy operator is typically a sum of 2-local non-commuting terms and the spectrum is continuous---the U(1)-constrained ensemble $\Phi_M$ serves as the asymptotic ensemble for systems with a single local scalar charge, regardless of the nature of the charge. This is because, in the limit of large subsystem dimension $d_A \gg 1$, non-universal microscopic details are washed away, and eigenstates effectively behave as random states subject to constraints of spatial locality and energy conservation. The correspondence between eigenstate statistics and $\Phi_M$ can be observed even in finite-size numerics, as shown below, provided the Hamiltonian parameters are chosen to maximize eigenstate entropy---what we refer to as `maximally chaotic' parameters. For these reasons, we will use the entanglement statistics of $\Phi_M$ as the reference distribution for eigenstates of quantum chaotic Hamiltonians.

\section{Ensembles emerging from U(1) Random Quantum Circuits}
\label{sec:U(1)}

\subsection{Model}

To illustrate the above results in a simple physical model, we first consider a random quantum circuit model~\cite{2023AnnRev_randomcircuits} comprised of a one-dimensional chain with $L$ sites and periodic boundary conditions, where each site contains a spin 1/2 degree of freedom. The time-evolution is constrained to conserve the total $z$ component of the magnetization, $\hat{S}_z = \sum_j \hat{Z}_j$. In particular, we use brickwork circuits with staggered layers of unitary gates with range $r=2$~\cite{2024PRB_U1circuits}. The range $r$ is the number of contiguous qubits each individual gate acts on, so that $\hat{U}_{j, j+1, \cdots ,j+(r-1)}$ acts on sites $(j, j+1, \cdots ,j+r-1)$. 
Each gate is a $d^r \times d^r$ block diagonal matrix, with $(r+1)$ blocks labeled by the total $z$ charge of the qubits on $r$ sites: $\sum_{i=j}^{j+r-1} \hat{Z}_i$.  The blocks have size $\binom{r}{n}$ with $n=0,1,...,r$.  In the $r=2$ case, the gates $\hat{U}_{j, j+1}$ have the structure:
\begin{equation}
    \hat{U}_{j,j+1} = {\tiny\begin{pmatrix} 
    \uparrow \uparrow & \uparrow\downarrow\quad \downarrow\uparrow & \downarrow\downarrow \\
    \begin{pmatrix}  \\   U(1) \\  \\  \end{pmatrix}  \\ 
    & \begin{pmatrix} & &   \\  & U(2) & \\   & &  \\ \end{pmatrix} & \\
    & &  \begin{pmatrix}   \\  U(1) \\  \\  \end{pmatrix}   \end{pmatrix}},
    \label{eq:U1unitary2site}
\end{equation}
comprised of  (i) a $1 \times 1$ block acting in the $\ket{\uparrow \uparrow}$ subspace, (ii) a $2 \times 2$ block acting in the $\ket{\uparrow\downarrow}$, $\ket{\downarrow \uparrow}$ subspace, and (iii) a $1 \times 1$ block acting in the $\ket{\downarrow \downarrow}$ subspace.

At time step $t$, we first apply a layer of unitary gates $\hat{U}_{\rm o}(t)$ acting on odd-even sites, followed by a second layer of unitary gates $\hat{U}_{\rm e}(t)$ acting on even-odd sites:
\be
\hat{U}_{\rm o}(t)  = \prod_{j=1}^{L-1} \hat{U}_{j,j+1}(t),\quad \hat{U}_{\rm e}(t) = \prod_{j=2}^{L} \hat{U}_{j,j+1}(t),
\ee
such that the total unitary at time $t$ is $\hat{U}(t) = \hat{U}_{\rm o}(t)\hat{U}_{\rm e}(t)$.
At each time step, each of the $\hat{U}_{j,j+1}(t)$ gates in Eq.~(\ref{eq:U1unitary2site}) are independently and randomly chosen. 
 
To obtain the late-time ensemble, we compute sequence of states $|\Psi_{t+1}\rangle = \hat{U}(t)|\Psi_t\rangle$ starting from the initial condition $|\Psi_0\rangle$. 
We first apply $t_{\rm eq} \gg L$ layers of unitary gates to reach equilibration, which we define as the timescale in which the average EE reaches its late-time value (within one standard deviation of the EE at late times). 
For $L=16$, we find that $t_{\rm eq}=100$ layers is sufficient to reach equilibration for all the initial conditions used. After equilibration, we sample quantum states at late times from a broad window of 1000 steps, with each sample separated by one layer of the quantum circuit. This ensures uniform sampling in the late-time regime while possibly allowing correlations between consecutive samples, which, however, are found to be negligible under quantum chaotic evolution (in Appendix~\ref{App:sampletosample}, we discuss the effects of state-to-state correlations).

\subsection{Initial Conditions}

We consider two qualitatively distinct product state initial conditions, the first describing spins aligned in the $x$ direction, and the second describing consecutive spins anti-aligned in the $z$ direction:
\begin{align}
\ket{{\rm FM}_x} & = \frac{1}{\sqrt{2^L}}\bigotimes_{j=1}^L \left( \ket{\uparrow}_j+\ket{\downarrow}_j \right),  
\label{eq:fmx} 
\\ 
\ket{{\rm AFM}_z} & = \ket{\uparrow\downarrow\uparrow\downarrow\ldots\downarrow}.
\label{eq:afmz}
\end{align}
 Both initial conditions have total magnetization $\langle \hat{S}_z\rangle = 0$. However, the FM$_x$ state has projection on all symmetry sectors $M$, with $p_M = d_M/2^L$, whereas AFM$_z$ has only projection in the largest $M=L/2$ sector. 
As such, FM$_x$ in Eq.~(\ref{eq:fmx}) satisfies the typicality in condition defined in Eq.~(\ref{eq:typical}), whereas the AFM$_z$ in Eq.~(\ref{eq:afmz}) is atypical and has zero magnetization variance.  

\begin{figure}
    \centering
    \includegraphics[scale=1]{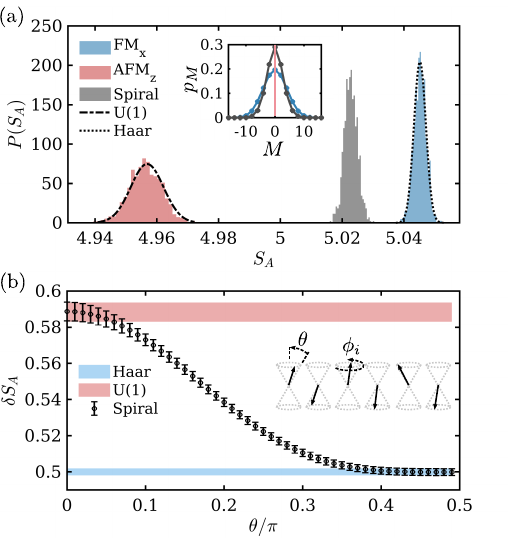}
    \caption{(a) Distribution of half-system entanglement entropy (EE) at late-times for a random quantum circuit with a U(1) conservation law. All the initial conditions considered have zero average magnetization $\langle \hat{S}_z \rangle = 0$
    but different magnetization variance $\delta S_z^2$, ranging from 0 (AFM$_z$) to $L/4$ (FM$_x$). Also shown is a generic EE distribution for a spin spiral state, Eq.~(\ref{eq:spiral}), with $\theta = \pi/4$. 
    The dotted line indicates the exact EE distribution for pure random states (Haar), and the dotted-dashed lines indicates the EE distribution of constrained random states [U(1)] derived in Ref.~\cite{2019PRD_BianchiDona}. 
    The inset shows the projection $p_M$ of the initial condition within symmetry sectors $M$ for each of the initial conditions. (b) Distribution of EE plotted relative to the maximum EE value ($\delta S_A = L_A\log 2 -S_A$) for spin spiral states, Eq.(\ref{eq:spiral}), as a function of the canting angle $\theta$.}
    \label{fig:U(1)}
\end{figure}

\subsection{Numerical Results}

Figure~\ref{fig:U(1)} shows the EE distribution at late times starting from the FM$_x$ and AFM$_z$ initial conditions. Regardless of the initial condition, both distributions exhibit volume-law entanglement entropy with leading order term equal to $L_A\log 2$. For this reason, we plot the distributions relative to the volume-law term $L_A\log d$, i.e., $\delta S_A = L_A\log 2 - S_A$,
thereby capturing the statistical behavior of the late-time ensemble at the level of ${\cal O}(1)$ corrections and statistical fluctuations. We find that both distributions are distinguishable at the level of subsystem entropies, as the their means are separated by a value much larger than their standard deviations. On the one hand, the FM$_x$ initial condition mixes all symmetry sectors according to the typicality condition, and the resulting EE distribution at late times is indistinguishable from the EE distribution of Haar random states (dotted lines). On the other hand, states obtained from the AFM$_z$ initial condition are restricted to the largest $M=L/2$ sector and agree exactly with the U(1) constrained distribution derived in Refs.~\cite{2019PRD_BianchiDona,2022PRX_eereview}.

We note that the constrained [U(1)] and unconstrained (Haar) distributions can be interpreted as the two {\it limiting} universal distributions for midspectrum states in generic quantum chaotic systems with a local conservation law. The former corresponds to the limiting case where states have zero variance of the conserved charge, while the latter corresponds to the opposite limit of large charge fluctuations, where states exhibit the same variance of the conserved charge as a pure random state. In particular, we can also generate any other EE distribution falling in-between these two limiting distributions, e.g., see the middle histogram in Fig.~\ref{fig:U(1)}(a). To show this, we use an `antiferromagnetic' spin spiral state initial condition defined as
\be
|\theta\rangle = \bigotimes_{j=1}^L \left(\cos\theta_j \ket{\uparrow}_j + e^{i\phi_j}\sin\theta_j\ket{\downarrow}_j\right),
\label{eq:spiral}
\ee
where the canting angle $\theta_j$ is equal to $\theta_j = \theta$ in odd sites, $\theta_j = \theta+\pi$ in even sites, and the azimuthal angle $\phi_j$ is $\phi_j = \frac{2\pi j}{L}$, see inset of Fig.~\ref{fig:U(1)}(b). By tuning the angle $\theta$, we go from an AFM$_z$ initial condition in Eq.~(\ref{eq:afmz}) with zero magnetization variance at $\theta=0$, to product states of spins spiraling in the $xy$ plane with magnetization variance that satisfies the typicality condition at $\theta = \pi/2$. 

The EE distribution as a function of $\theta$ is shown in Fig.~\ref{fig:U(1)}(b). Each datapoint represents the mean EE of the distribution, the bars indicate its standard deviation, and the EE is plotted relative to its maximum value ($\delta S_A = L_A\log 2 - S_A$). As anticipated, we find that the EE distribution at late times can fall anywhere in between that of $\Phi_{\rm Haar}$ or $\Phi_{U(1)}$ by tuning the value of $\theta$ between $0 \le \theta \le \pi/2$. In the next section, we show that the same behavior applies to more general situations describing local Hamiltonian systems, where the charge can no longer be decomposed as a sum of local commuting terms. 

\section{Ensembles emerging from chaotic Hamiltonian dynamics}
\label{sec:Ham}

\subsection{Model}

We now consider the mixed field Ising model (MFIM), 
\begin{equation}
    {\hat H} = \sum_{j} \hat{Z}_j\hat{Z}_{j+1} + g \hat{X}_j +h \hat{Z}_j ,
    \label{eq:Ham}
\end{equation}
which has been widely studied as a canonical model of strongly chaotic dynamics~\cite{2011PRL_banulscirachastings,2013PRL_Huseentanglementgrowth,2019PRX_lanczos,2024PRX_quantifyingchaos}. Here $\hat{X}_j, \hat{Y}_j, \hat{Z}_j$ denote the Pauli matrices, $g$ is the transverse field, and $h$ is the longitudinal field. The Hamiltonian (\ref{eq:Ham}) also has multiple point symmetries, which we explicitly break. In particular, we use open boundary conditions to break translation symmetry, and we include boundary fields $h_{1}=\hat{Z}_1/4$ and $h_{L}=-\hat{Z}_L/4$ to break inversion symmetry. Of special interest in this work are the model parameters $(g,h) = (1.1,0.35)$, which we have identified as the `most chaotic' parameters in terms of eigenstate randomness~\cite{2024PRX_quantifyingchaos}, and are also close to the values used in many other works~\cite{2011PRL_banulscirachastings,2019PRX_lanczos}. 

To obtain the late-time ensemble, we first evolve the initial state $|\Psi_0\rangle$ (defined below) for long times $t_{\rm eq} \gg 1$ until the EE equilibrates. As before, we define the thermalization time as the time in which the average EE reaches its late-time value (within one standard devition of the EE at late times). For all the system sizes $10 \le L \le 16 $ and initial conditions considered in this work, we find that $t_{\rm eq}=200$ is sufficient for the EE to equilibrate. After equilibration, we draw late-time quantum states from a broad temporal window of size $\Delta t = 100$. As in the U(1) case discussed in Sec.~\ref{sec:U(1)}, we sample quantum states {uniformly} within this window, possibly allowing for correlations between the reduced density matrix of subsequent samples. However, we find that these sample-to-sample correlations become negligible in ${\cal O}(1)$ timescales, thus EE datapoints can be assumed to be effectively uncorrelated (see Appendix~\ref{App:sampletosample}).

\begin{figure}
    \centering
    \includegraphics[scale=1.0]{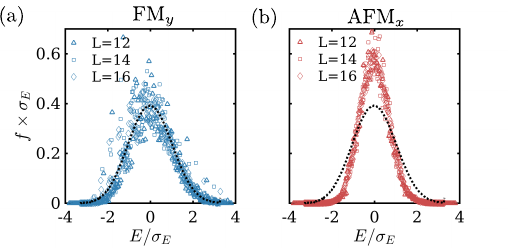}
    \caption{Projection of the (a) FM$_y$ and (b) AFM$_x$ states on the eigenstate basis, $f(E) = \sum_n |\langle n | \Psi_0\rangle|^2\delta(E-\varepsilon_n)$, of the mixed field Ising model (MFIM). In both panels, datapoints are shown for system sizes $L=12$ (triangle), $L=14$ (squares), and $L=16$ (diamonds). Also shown with dashed lines is the average projection of pure random states on the eigenstate basis (c.f., Fig.~\ref{fig:schematics}).}
    \label{fig:Haminitial}
\end{figure}

\subsection{Initial Conditions}

Similarly to the U(1) case, we consider two classes of midspectrum (or `infinite temperature') product state initial conditions that have distinct late-time behavior. First, we consider the $|{\rm FM}_y\rangle$ product state comprised of spins aligned in the $y$-axis. Second, we consider the AFM$_x$ initial comprised of spins anti-aligned in the $x$ direction. Both initial conditions have zero energy, $\langle \Psi_0 | {\hat H} | \Psi_0 \rangle = 0$, but different energy variance. On the one hand, the FM$_y$ initial condition has energy variance $\sigma_E^2 = \langle \Psi_0 | {\hat H}^2 | \Psi_0 \rangle - \langle \Psi_0 | {\hat H}| \Psi_0 \rangle^2 = L(1+g^2+h^2)$, which agrees with that of Haar random states, therefore FM$_y$ is typical. On the other hand, the AFM$_x$ initial condition has energy variance $\sigma_E^2 = \langle \Psi_0 | {\hat H}^2 | \Psi_0 \rangle - \langle \Psi_0 | {\hat H}| \Psi_0 \rangle^2 = L(1+h^2)$, which is smaller than that of Haar random states, and therefore AFM$_x$ is atypical. More explicitly, the spreading of both initial conditions on the eigenstate basis of ${\hat H}$ is shown in Fig.~\ref{fig:Haminitial} for $L = 12, 14, 16$ (c.f., Fig.\ref{fig:schematics}).

For comparison, we also construct the analogue of the microcanonical ensemble $\Phi_M$ comprised of states with zero energy variance. More precisely, we define the microcanonical {\it eigenstate} ensemble $\Phi_E$ as $\Phi_E = \{ |n\rangle,\, |E_n| < \epsilon\}$, with eigenstate ${\hat H}\ket{n} = E_n\ket{n}$ drawn from a small energy window of size $\epsilon \ll 1$ around the middle of the spectrum. To numerically generate the microcanonical ensemble, we first diagonalize $\hat{H}$ in Eq.~\eqref{eq:Ham}, and then sample $N_{\rm mid}$ eigenstates in the middle of the spectrum to obtain $\Phi_E$. We use $N_{\rm mid} = 100,300,600,1000$ midspectrum eigenstates for system sizes $L = 10,12,14,16$, respectively.

\begin{figure}
    \centering
    \includegraphics[scale=1.0]{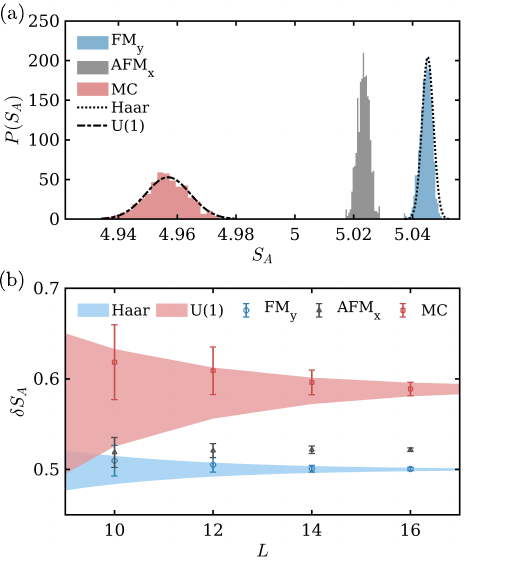}
    \caption{(a) Distribution of EE at late times for the MFIM as a function of initial conditions. Also shown is the microcanonical distribution of EE for midspectrum eigenstates (MC). The dotted lines indicate the EE distribution of Haar random states, and the dotted-dashed lines indicate the EE distribution for states constrained to the $M=0$ magnetic sector. (b) Finite-size scaling of the EE distributions plotted relative to the maximum EE value ($\delta S_A = L_A\log 2-S_A$) for different initial conditions. The dots indicated the average EE value, and the bars indicated their standard deviation. The shaded areas indicates the regions limited by $S_A = \mu_{\rm H}\pm\sigma_{\rm H}$ (blue) and $S_A = \mu_M\pm\sigma_M$ (red) for the EE distribution of Haar and constrained random states, respectively.}
    \label{fig:Ham}
\end{figure}

\subsection{Numerical Results}

Figure~\ref{fig:Ham}(a) shows the late-time EE distribution for the FM$_y$ (typical) and AFM$_x$ (atypical) initial conditions, and the microcanonical eigenstate EE distribution of midspectrum eigenstates (MC) for the MFIM with $(g,h) = (1.1,0.35)$ and $L=16$. We first emphasize the striking similarity between Figs.~\ref{fig:Ham} and \ref{fig:U(1)}. On the one hand, the FM$_y$ initial condition, which projects onto the eigenstate basis uniformly across eigenstates [Fig.~\ref{fig:Haminitial}(a)], exhibits an EE distribution that is indistinguishable from that of Haar random states, up to corrections that are exponentially small in system size. On the other hand, the AFM$_x$ initial condition, with smaller-than-typical energy variance [Fig.~\ref{fig:Haminitial}(b)], leads to an EE distribution that is distinguishable from the Haar ensemble by several standard deviations. In other words, the mean EE exhibits ${\cal O}(1)$ corrections to the Page entropy that are much larger than the EE fluctuations. 
In the limiting case of quantum states with zero energy variance, the microcanonical eigenstate ensemble $\Phi_E$ exhibits an EE distributions that closely matches that of random state ensembles constrained by a U(1) conservation law, as noted in Ref.~\cite{2024PRX_quantifyingchaos}.\footnote{Because the Hamiltonian (\ref{eq:Ham}) has time-reversal symmetry, we sample $\Phi_{E}$ using real-valued random states with constraints. This does not modify the mean EE, but its standard deviation increases by a factor of $\sqrt{2}$~\cite{2010JPA_eermt,2011JPA_rmtee,2024PRX_quantifyingchaos}. This result is true both for  constrained~\cite{2024PRX_quantifyingchaos} and unconstrained ensembles~\cite{2010JPA_eermt,2011JPA_rmtee}.} 
In particular, the mean EE of Hamiltonian eigenstates has an ${\cal O}(1)$ correction to the Page entropy equal to $\frac{1}{2}[f+\log(1-f)]$ (within the precision of finite-sized numerics), and the EE fluctuations scale as $\sigma_E \sim \sqrt{L}/d^L$, as discussed in Sec.~\ref{sec:EEU1}.

These ${\cal O}(1)$ corrections to the Page entropy, along with the larger fluctuations, indicate that eigenstates—and more generally, atypical states—encode more information than Haar-random states. This additional information was shown to arise from the constraints of spatial locality and energy conservation \cite{Huang2021,2024PRX_quantifyingchaos}. This additional information is also what enables the full reconstruction of the Hamiltonian from a single eigenstate \cite{2018PRX_Grover,2019Q_ranard}. The behavior of $\Phi_E$ differs markedly from that of the late-time ensemble of typical initial states: at late times, dynamical evolution effectively erases spatial locality, leading the ensemble to exhibit Haar-like behavior in its finite statistical moments.

\begin{figure*}
    \centering
    \includegraphics[scale=1.0]{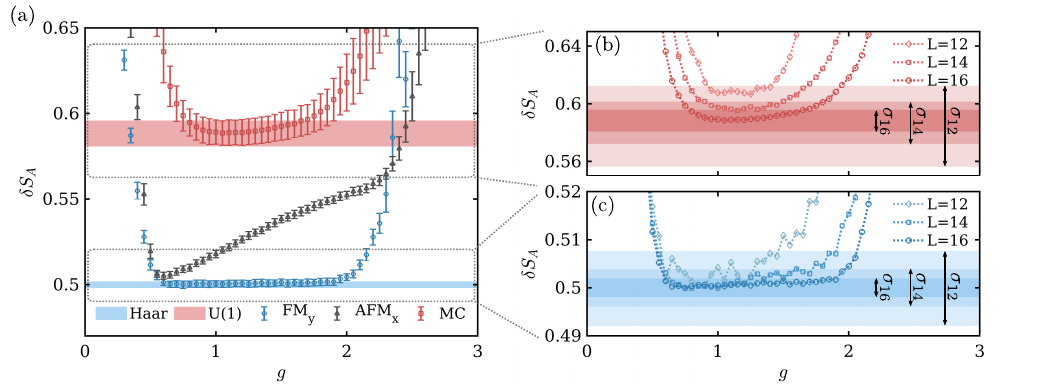}
    \vspace{-3mm}
    \caption{(a) Distribution of EE at late-times for the MFIM as a function of the transverse field $g$ and initial conditions (FM$_y$ and AFM$_x$). Also shown is the EE distribution of the microcanonical eigenstate ensemble (MC). All values are plotted relative to the maximum EE ($\delta S_A = L_A\log 2 - S_A$). The symbols indicate the average EE value for each value of $g$, and the bars indicate their standard deviation. The shaded areas indicates the regions limited by $S_A = \mu_{\rm H}\pm\sigma_{\rm H}$ (blue) and $S_A = \mu_M\pm\sigma_M$ (red) for the EE distribution of Haar and constrained random states, respectively. (b,c) Zoom of the dotted regions in panel around (b) showing the finite size scaling of the EE distribution for (b) eigenstates and (c) late-time ensembles for the FM$_y$ initial condition, plotted for system sizes $L=12,14,16$. The width $\sigma_L$ of the shaded regions indicate the standard deviation of the reference EE distribution for the constrained and unconstrained random states, for each system size.}
    \vspace{-4mm}
    \label{fig:Hamg}
\end{figure*}
 
Figure~\ref{fig:Ham}(b) shows the finite-size scaling behavior of panel (a), showing that the same behavior described above persists in the thermodynamic limit. Each datapoint represents the mean EE at late times for different system sizes $L$ and initial conditions, and the bars represent the standard deviation of the EE distribution. For comparison, the shaded areas indicates the regions limited by $S_A = \mu_{\rm H}\pm\sigma_{\rm H}$ [Eq.~(\ref{eq:Haarfirstmoment})] and $S_A = \mu_M\pm\sigma_M$ [Eq.~(\ref{eq:U1firstmoment})] for the EE distribution of states within $\Phi_{\rm Haar}$ and $\Phi_{M}$ ensembles, respectively. We find excellent convergence of all the distribution to their expected universal RMT behavior for the typical and the MC distributions. In addition, the atypical initial conditions remain distinguishable from Haar and microcanonical distributions for all system sizes.

So far, our results focused on the late-time behavior of typical and atypical initial states for maximally chaotic Hamiltonians. In particular, the numerical results in Fig.~\ref{fig:Ham} were obtained for the most chaotic parameters $(g,h) = (1.1,0.35)$ of the MFIM. We now show that the main conclusions of this work are generic, i.e., the late-time ensemble of typical initial states are indistinguishable from Haar random states regardless of Hamiltonian details, so long as the Hamiltonian remains in the quantum chaotic regime. With this goal in mind, we use the same initial conditions (FM$_{y}$ and AFM$_x$) in the MFIM and tune the value of $g$ away from the most chaotic parameters. As shown in Fig.~\ref{fig:Hamg}(a), the late-time EE distribution for the typical (FM$_y$) initial condition approaches the EE distribution of Haar random states for a broad range of $g$ values. In addition, the finite-scaling of the results in Fig.~\ref{fig:Hamg}(c) shows that the range of $g$'s in which the late-time EE distribution approaches the Haar ensemble increases with system size. This suggests that, in the thermodynamic limit, the late-time distribution of quantum states becomes indistinguishable from Haar random states for all values of $g$ so long as the Hamiltonian is quantum chaotic. 

The behavior is qualitatively distinct for atypical initial states (AFM$_x$), as their late-time behavior exhibit a strong sensitivity to the Hamiltonian parameters, as shown in Fig.~\ref{fig:Hamg}(a). This behavior is analogous to that shown in Fig.~\ref{fig:U(1)}(b) for spin spiral states, where we found that details of the initial condition modify the late-time behavior of quantum states, and such finer aspects can be distinguished through ${\cal O}(1)$ corrections to the EE. 

Finally, we examine the behavior of the microcanonical eigenstate ensemble, $\Phi_E$, across different model parameters. First, we find that the eigenstate EE exhibits ${\cal O}(1)$ corrections relative to the Page entropy, independent of the model parameters. This appears as a vertical shift in the EE distributions shown in Fig.\ref{fig:Hamg}(a) for all values of $g$, which persists in the thermodynamic limit, as seen in Fig.\ref{fig:Hamg}(b). As discussed above, an EE below the Page value indicates that eigenstates contain more structure and information than Haar-random states. Second, Ref.~\cite{2024PRX_quantifyingchaos} found that the microcanonical eigenstate ensemble approaches the behavior of U(1)-constrained random states near the Hamiltonian parameters $(g,h)=(1.1,0.35)$ even for relatively small system sizes. This behavior is shown in Fig.\ref{fig:Hamg}(b), where the EE statistics of eigenstates is found to be within one standard deviation of the reference random state distribution even for relatively small system sizes $L=12$. This observation led to dubbing these parameters ‘maximally chaotic,’ as the minimum distance between the eigenstate and reference distributions is reached at the same point in Hamiltonian parameter space, independently of $L$. In contrast, the late-time ensemble does not exhibit the same behavior. As discussed above, we instead find excellent agreement with Haar-like behavior across a broader range of model parameters.
This trend is shown in Fig.\ref{fig:Hamg}(c) for the mean EE, while the comparison for the second moments is presented in Appendix~\ref{App:gfluctuations}. These findings suggest that, regardless of the specific model parameters, correlations induced by spatial locality are typically erased by quantum chaotic evolution at late times.

In the discussion above, we presented all numerical data in terms of the distributions of half-system von Neumann entanglement entropy (EE), as it serves as a highly sensitive probe of quantum state randomness. However, as argued earlier, the conclusions of this work do not depend on the choice of subsystem observable or the subsystem size. In Appendix~\ref{App:Renyi}, we explicitly demonstrate this by extending our numerical results to smaller subsystem sizes and to the second Renyi entropy.

\section{Summary and Discussion}
\label{sec:discussion}

Our work has shown that quantum states undergoing quantum chaotic dynamics in the presence of symmetries can exhibit rich classes of dynamical behaviors at late times, even when initial states are drawn from the middle of the spectrum. These rich behaviors emerge when studying higher statistical moments of the quantum state ensembles, particularly the universal behavior of state-to-state fluctuations at late times beyond the widely-studied coarse-grained behaviors, such as the volume-law behavior of the entanglement entropy. 

In particular, our work has shown that typical product state initial conditions in quantum chaotic systems effectively mix symmetry sectors in such a way that finite $k$-moments of the late-time ensemble exactly agree with Haar random states in the thermodynamic limit. This means that no measurement can be done to tell quantum states drawn from dynamics apart from Haar random states. For finite-sized systems, it is exponentially hard to tell both ensembles apart. This result is quite striking, as the ensemble of quantum states at late-time do not explore the entire Hilbert space when evolution is constrained by symmetries. In contrast, initial states with negligible variance of the conserved charge also acquire universal behavior which can be described by pure random state ensembles constrained within a given symmetry sector. As a result, one can define two limiting regimes controlled by the variance of the conserved quantity: states with variance equal to that of Haar random states give rise to effectively Haar random states, and states with zero variance give rise to constrained random state ensembles. Initial states that lie in between both limiting regimes exhibit non-universal late-time dynamics with qualitative features equal to those of Haar random states, such as volume-law EE, but with distinct features at the level of ${\cal O}(1)$ corrections and fluctuations.

In addition to characterizing the universal structure of late-time ensembles in quantum chaotic systems, we also emphasize several consequences of our work. The first relates to our understanding of quantum chaos. The widely-accepted definition of quantum chaos is rooted in the RMT behavior of eigensystem properties~\cite{Atas2013,Atas2013a,2007PRB_ratiofactor,2018PRL_exactsff,Kos2018,2018PRX_qcdiagrammatics,Roy2020,Murthy2019,2019PRE_lugrover,2023PRE_averageEE, 2022PRX_eereview}. As we argued above, this definition only captures the average behavior of quantum states and applies generically to any Hamiltonian away from integrable limits. When looking at higher statistical moments, our work finds qualitatively distinct differences between the eigenstate ensemble and the late-time ensemble. On the one hand, typical initial states evolved under quantum chaotic Hamiltonian dynamics exhibit the same universal statistical properties---as captured by finite statistical moments---as the Haar ensemble, regardless of microscopic details [Fig.~\ref{fig:Hamg}]. On the other hand, the eigenstate ensemble exhibits ${\cal O}(1)$ corrections to the Page entropy and larger statistical fluctuations. This shows that eigenstates encode more information than quantum states at late times. The key distinction between eigenstate and temporal ensembles is the role of spatial locality: whereas spatial locality is imprinted in the structure of eigenstates, spatial locality is washed away during dynamics at late times. The higher statistical moments of the quantum state ensembles can tell such differences apart. An interesting direction for future work is creating sharper definitions of quantum chaos that account for higher statistical moments of quantum states, both at the level of eigenstates and temporal ensembles.

A second consequence of our work relates to randomization of quantum information in quantum devices. Modern experimental platforms operate under a large number of constraints, including spatial locality (gates operate between neighboring qubits) and a restricted subset of local gates~\cite{2019Nature_quantumsuppremacy,2024Science_googlekpz,2021Science_googlescrambling,2017Nature_51atomsimulator,2021Nature_Rydberg256,2021Nature_RydbertAFM,2017Nature_monroe,2021RMP_trappedtions}. This raises questions about how much randomness a quantum device can truly generate~\cite{2019Nature_quantumsuppremacy,2021Science_googlescrambling,2023Nature_emergentdesign}. Our results suggest that even when quantum evolution is constrained by locality and simple symmetries—preventing the generation of all possible states—one can still obtain a late-time ensemble that is effectively indistinguishable from the Haar distribution, provided the initial condition is appropriately chosen. Generalizing our results to more complex constraints is an interesting direction for future research.

A third consequence relates to the existence of atypical midspectrum initial conditions that exhibit smaller-than-typical EE at late times. Although midspectrum states with smaller-than-typical entanglement are reminiscent of many-body quantum scars~\cite{2018NP_quantumscars,2021NatPhys_scarreview}, we emphasize that quantum scars have anomalously small entanglement entropy due to the absence of volume law behavior, whereas the atypical states discussed here still have volume law behavior but smaller-than-typical EE at the level of $\mathcal{O}(1)$ corrections and statistical fluctuations.  Also, quantum scars are fine-tuned and require special phase space structure of the Hamiltonian~\cite{2023AnnRev_scars}, possibly proximity to integrability~\cite{2019PRB_scarskhemani}, whereas the atypical states discussed here are robust and ubiquituous, persisting irrespective of the Hamiltonian details (Fig.~\ref{fig:Hamg}). Understanding in more detail the dynamic and thermalizing properties of these atypical states is another interesting direction for future research.

\section*{Acknowledgements}

We thank Wen Wei Ho, Daniel Mark, Soonwon Choi, and Shenglong Xu for insightful comments, and Vedika Khemani for previous collaborations. JFRN acknowledges the hospitality of the Aspen Center for Physics, which is supported by National Science Foundation grant PHY-2210452, and a grant from the Alfred P. Sloan Foundation (G-2024-22395). The numerical simulations in this work were conducted with the advanced computing resources provided by Texas A\&M High Performance Research Computing.

\appendix
\renewcommand{\thefigure}{A\arabic{figure}}
\setcounter{equation}{0}
\setcounter{figure}{0}

\section{Trace distance between Haar random states and typical random states}

To compute the trace distance between the second moment of the ensemble of Haar random states, Eq.~(\ref{eq:Haarsecondmoment}), and the ensemble of typical random states, Eqs.~(\ref{eq:U1secondmoment1}) and \eqref{eq:U1secondmoment2}, we first note that the matrix $\delta\rho^{(k=2)}$ is a sparse matrix that has non-zero off-diagonal entries only for permutations of the $m_1\neq m_2$ indeces, {i.e.}, $\delta \rho_{m_1m_2,m_2m_1}^{(k=2)}\neq 0$. To diagonalize this sparse matrix, we use the basis $|M,\alpha\rangle$ of the $Z$ operator used in Eqs.~\eqref{eq:U1secondmoment1}--\eqref{eq:U1secondmoment2}. 
We consider three distinct cases. 

(i) When $M_1 \neq M_2$, there is a total of $ \sum_{M_1\neq M_2} D_{M_1}D_{M_2}$ terms that contribute to the trace distance. Each state $(M_1\alpha_1,M_2\alpha_2)$ is coupled to the state obtained by permuting its indices, $(M_2\alpha_2,M_1\alpha_1)$, and the eigenvalues of the two-by-two matrix describing this subspace are 0 and $\frac{1}{D^2}-\frac{1}{D^2(1+1/D)}$. We also note that $\sum_{M_1\neq M_2} D_{M_1}D_{M_2} = \sum_{M_1M_2}D_{M_1}D_{M_2} - \sum_{M}D_{M}^2 = D^2-\sum_MD_M^2$.

(ii) When $M_1 = M_2 = M$ and $\alpha_1\neq\alpha_2$, for {\it each} symmetry sector $M$ there is a total of $D_M(D_M-1)$ terms that contribute to the trace distance. Each state $(M\alpha_1,M\alpha_2)$ is only coupled to the state obtained by permuting its indeces, $(M\alpha_2,M\alpha_1)$, and the eigenvalues of the two-by-two matrix describing this subspace are 0 and $\frac{1}{D^2(1+1/D)}-\frac{1}{D^2(1+1/D_M)}$. 

(iii) When $M_1 = M_2 = M$ and $\alpha_1=\alpha_2$, there is a total of $D_M$ terms that contribute to the trace distance in {\it each} symmetry sector $M$. Each 
term contributes $\frac{1}{D^2(1+1/D_M)}-\frac{1}{D^2(1+1/D)}$ to the trace distance. 

Combining all three contributions (i)-(iii), the trace distance is given by 
\begin{align}
\Delta_{k=2} & = \frac{1}{2}\left(1-\frac{1}{D^2}\sum_MD_M^2\right)\left(1-\frac{1}{1+1/D}\right) \nonumber\\
 & + \sum_M \frac{D_M(D_M-1)}{2D^2}\left( \frac{1}{1+1/D} - \frac{1}{1+1/D_M}\right) \nonumber\\
 & + \sum_M \frac{D_M}{D^2} \left(\frac{1}{1+1/D_M} - \frac{1
}{1+1/D}\right)
 \label{eq:delta2exact}.
\end{align}
We next expand Eq.~\eqref{eq:delta2exact} up to order ${\cal O}(1/D^2)$. The term in the first line of the right-hand-side is 
\begin{align}
\left( 1-\frac{1}{D^2}\sum_MD_M^2 \right)\left(1-\frac{1}{1+1/D} \right) \approx  \frac{1}{D}\left( 1-\frac{1}{\sqrt{\pi L}}\right),
\label{seq:delta1}
\end{align}
where we used the equality $\sum_M D_M^2 = \sum_{M=0}^L {L\choose M}^2 = {2L \choose L}$, combined with Stirling's approximation ${2L \choose L} \approx \frac{2^{2L}}{\sqrt{\pi L}} = \frac{D^2}{\sqrt{\pi L}}$. Second, we combine all the terms that have denominator $\frac{1}{1+1/D}$ in the second and third lines of Eq.~\eqref{eq:delta2exact}. This gives 
\begin{align} 
\sum_M \frac{D_M(D_M-3)}{D^2(1+1/D)} \approx & \left(\frac{1}{\sqrt{\pi L}} - \frac{3}{D}\right)\left(1-\frac{1}{D}\right) \nonumber \\ 
= & \left(\frac{1}{\sqrt{\pi L}} - \frac{3}{D}-\frac{1}{D\sqrt{\pi L}}\right),
\label{seq:delta2}
\end{align}
where we used again the approximation $\sum_M D_M^2 \approx \frac{D^2}{\sqrt{\pi L}}$.
Third, we combine all the terms that have denominator $\frac{1}{D^2(1+D_M^{-1})}$ in the second and third lines of Eq.~\eqref{eq:delta2exact}. This gives 
\begin{align}
\sum_M \frac{D_M(D_M-3)}{D^2(1+D_M^{-1})} = & \sum_M \frac{D_M(D_M-3)}{D^2}\left( 1 - \frac{1}{D_M}+\ldots\right) \nonumber\\ 
\approx & \sum_M\frac{D_M^2-4D_M}{D^2} = \left(\frac{1}{\sqrt{\pi L}} - 4D\right)
\label{seq:delta3}
\end{align}
By summing all the terms in Eqs.~\eqref{seq:delta1}--\eqref{seq:delta3}, we obtain Eq.~\eqref{eq:Delta2} in the main text. 

\begin{figure}
    \centering
    \includegraphics[width=\columnwidth]{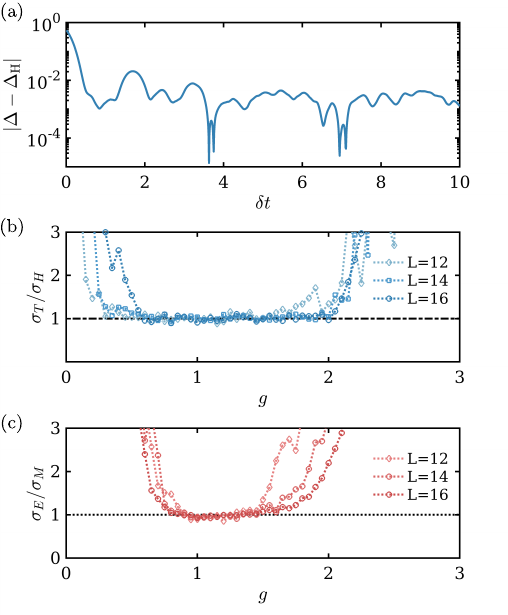}
    \vspace{-6mm}
    \caption{(a) Sample-to-sample correlations as measured through the trace distance $\Delta$ between the reduced density matrix at $t=200$ and time $t = 200+\delta t$ for the FM$_y$ initial condition. The data is shown relative to typical trace distance $\Delta_{\rm H}$ between two pure random states drawn from the Haar ensemble. The numerical data, shown for the MFIM with $L=16$ spins, indicates that the reduced density matrices (from which EE is computed) become effectively random after ${\cal O}(1)$ timesteps. 
    (b,c) Second moments of the (b) late-time and (c) eigenstate ensembles, $\sigma_T$ and $\sigma_E$, respectively, shown relative to their corresponding reference value, namely, $\sigma_H$ of Haar random states and $\sigma_M$ of Haar random states constrained to a microcanonical symmetry sector.}
    \label{fig:tCorr}
\end{figure}

\section{Sample-to-sample correlations in the late-time ensemble}
\label{App:sampletosample}

The {\it late-time} ensemble is constructed from samples drawn from the dynamics at arbitrarily long times. When drawing a finite number of samples from this infinitely broad temporal window, the probability of two samples being separated by a short time is negligible. Thus, the samples typically exhibit negligible sample-to-sample correlations. 

When producing the numerical data, we are necessarily limited to sampling within a finite temporal window. However, we carefully chose a sufficiently large temporal window that closely approximates the late-time ensemble. While it is possible for some samples to be close in time, their contribution to statistical correlations should be minimal. Typically, one expects that over timescales of ${\cal O}(L)$, samples effectively become uncorrelated. However, evidence from quantum circuits suggests that decorrelation may even occur faster~\cite{2024arxiv_huangcircuitdepth}.

Having these considerations in mind, for the Hamiltonian data we first evolve the initial state up to time $T=200$, which is an order of magnitude larger than $L$, ensuring that the first two moments of the entanglement entropy (EE) have equilibrated (i.e., reached their late-time value). We then draw 1000 samples uniformly from a large temporal window of size $\Delta T = 100$. While samples that are very close in time within this finite window exhibit sample-to-sample correlations---such as those shown in Fig. \ref{fig:tCorr} via the trace distance of unequal-time reduced density matrices---these correlations do not significantly impact the final results. For instance, doubling the temporal window (thus increasing the average time separation between samples to twice its value) results in statistically insignificant changes in the EE distribution, with the mean changing by only 0.004\%. 

\begin{figure}
    \centering
    \includegraphics[width=\columnwidth]{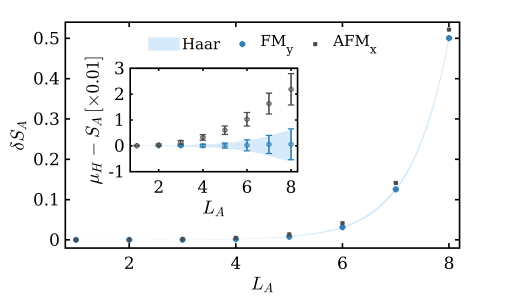}
    \vspace{-3mm}
    \caption{Distribution of EE for the MFIM as a function of subsystem size $L_A$ and initial conditions for $L=16$. The data in the main panel is shown relative to the maximum EE value, i.e. $\delta S_A = L_A\log 2 - S_A$. The error bars of the datapoints in the main panel are smaller than the symbol size. 
    The narrow shaded area indicates the regions limited by $S_A = \mu_{\rm H}\pm\sigma_{\rm H}$ (blue) for the EE distribution of Haar random states. The inset shows the EE distribution of the late ensemble relative to the exact Page entropy, thereby highlighting differences between distributions on the scale of statistical fluctuations. The results show excellent agreement for the typical initial condition FM$_y$, and statistically significant deviations for the atypical initial condition AFM$_x$.}
    \label{fig:subsys}
\end{figure}

\section{Fluctutations of EE across model parameters}
\label{App:gfluctuations}

In this section, we extend the results of Fig.\ref{fig:Hamg}(b) and (c) to examine the second moments of EE fluctuations across ensembles, specifically for the late-time ensemble and the microcanonical eigenstate ensemble. These results are presented in Fig.\ref{fig:tCorr}(b) and (c), respectively. Similar to the behavior of the average EE, we find that the EE fluctuations of the late-time ensemble of typical initial states agrees with the behavior of the Haar ensemble for a wide range of model parameters. In contrast, the eigenstate ensemble agrees remarkably well with the statistical behavior of contrained random states only in the proximity of the maximally chaotic parameters $(g,h) = (1.1,0.35)$. This holds even for moderately small system sizes $L =12$. These behaviors are consistent with the results found for the mean EE in the main text. 

\begin{figure}[b]
    \centering
    \includegraphics[width=\columnwidth]{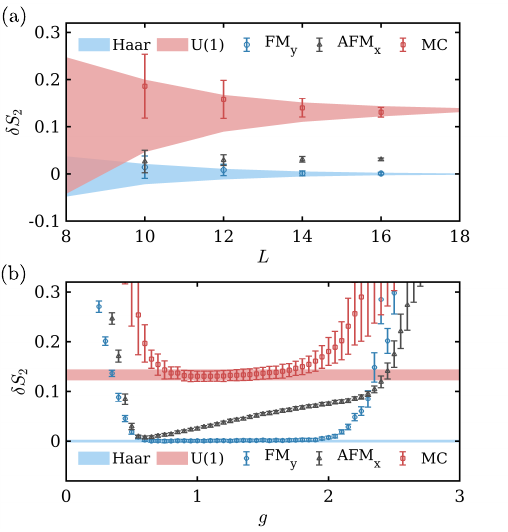}
    \vspace{-3mm}
    \caption{(a) Finite-size scaling of the distribution of second Renyi entropy, $S_2$, plotted relative to the maximum $S_2$ value ($\delta S_2 = L_A\log 2-\log 2 -S_2$) for different initial conditions (FM$_y$ and AFM$_x$). Also shown is the $S_2$ distribution of the microcanonical eigenstate ensemble (MC). The dots indicated the average $S_2$ value, and the bars indicated their standard deviation. The shaded areas indicates the mean $\pm$ standard deviation of the $S_2$ distribution for Haar (blue) and constrained (red) random states, respectively. (b) Distribution of $S_2$ at late-times for the MFIM as a function of the transverse field $g$ and initial conditions. }
    \label{fig:Renyi}
\end{figure}

\section{Complementary data for small subsystems and Renyi entropy}
\label{App:Renyi}

In the main text, we primarily show numerical data on half-system EE distributions at late times as the proxy of equilibration. As we argue at the beginning of Sec.~\ref{sec:EE}, because the EE is a nonlinear function of $\rho_A$, it is an an extremely sensitive sensitive probe of quantum state randomness; it is also a well-studied metric with known analytical results in various regimes of interest. However, we emphasize that none of our results depend on subsystem size, nor on the choice of subsystem observable. 

First, if quantum states at late times are indistinguishable at the level of half-systems---meaning their EE distribution matches that of Haar-random states---then it is reasonable to expect the same to hold for smaller subsystems, which are known to thermalize more effectively. This is a well-established fact in the quantum thermalization community. We numerically confirm this intuition for the MFIM in Fig.~\ref{fig:subsys}. Specifically, for the typical initial condition FM$_y$, we find excellent agreement with the Haar-random EE distribution for all subsystem sizes, not only in terms of the average behavior (main panel) but also in the fluctuations (inset). In contrast, for the atypical initial condition AFM$_x$, we observe significant deviations from Haar-random behavior on the scale of EE fluctuations. This occurs despite the expectation that smaller subsystems—being more prone to thermalization—should exhibit better agreement.

Second, while the von Neumann EE is a standard quantity in the literature, our results could just as easily be formulated in terms of other observables that are easier to measure experimentally, such as the second Renyi entropy $S_2$. In Fig.\ref{fig:Renyi}, we present the counterparts of Fig.\ref{fig:Ham}(b) and Fig.~\ref{fig:Hamg}(a), replacing the von Neumann EE with the Renyi entropy. We find exactly the same qualitative behavior, demonstrating that the conclusions of the main text extend directly to observables beyond von Neumann entropy.

\end{document}